# BOSE FLUIDS ABOVE T$_C$: INCOMPRESSIBLE VORTEX FLUIDS AND "SUPERSOLIDITY"

P W Anderson  Princeton University

Abstract:  **This paper emphasizes that non-linear rotational or diamagnetic susceptibility is characteristic of Bose fluids above their superfluid Tc's.  For sufficiently slow rotation or, for superconductors, weak B-fields this amounts to an incompressible response to vorticity. The cause is that there are terms missing in the conventionally accepted model Hamiltonian for quantized vortices in the Bose fluid.  The resulting susceptibility can account for recent observations of Chan et al on solid He, and Ong et al on cuprate superconductors.**

The recent experiments of Chan and students[1] have been generally interpreted as indicating that solid He is showing supersolid behavior.  Chan himself describes his observations as demonstrating what Leggett [2] described as "non-classical rotational inertia" (NCRI), that is the moment of inertia for low angular velocity is not that of a rigid rotor of the same density and dimensions. We prefer the designation "non-linear rotational susceptibility" (NLRS) which means much the same, namely that the moment of inertia is a function of the rotational velocity returning to the classical value at high values, while it indicates that I, the moment of inertia, has the character of a susceptibility to rigid rotation at angular velocity $\omega$:

$$I = \partial L / \partial \omega = \partial^2 F /(\partial \omega)^2 \quad [1]$$

Numerous experiments are unable to find true supercurrents flowing in solid He and the simplest explanation of these observations is that the phenomenon observed is simply NLRS and not superfluidity, at least under the conditions tested, which leaves the question of supersolidity open.

I observe that there is, in the experiments of Ong et al[3] on superconductors above Tc, a phenomenon of "non-linear diamagnetic susceptibility" (NLMS) in the absence of true superconductivity. The diamagnetic susceptibility, which is the response $\partial^2 F/(\partial B)^2$ to the vorticity in the electron gas induced by a magnetic field B, is relatively large and non-linear at field scales which are relatively low—in some cases it is even divergent at low B. On the other hand, the resistivity is finite and perfectly linear.

In the phase space region which Ong is investigating, the electrons are thought to be fairly strongly paired, superconductivity having been destroyed only by phase fluctuations of the order parameter, as demonstrated in 1993 by Salomon[4]. Therefore it is reasonable to think of the currents as being predominantly carried by paired electrons, i e bosons. If there is a finite local pair amplitude above Tc, the pair wave function will have a time- and space-varying phase $\Phi$ and the current will be proportional to $\nabla\Phi$ and conserved. If so $\Phi$ will be completely determined by a network of vortex lines—in 3D, mostly vortex loops. Thus it is appropriate to describe this phase as a vortex fluid.[5]

In solid He the currents, whether flowing in some percolating network of defects, as many believe, or intrinsic to an incommensurate solid, are necessarily bosonic, and may also be describable by a local time- and space- varying phase. In this case it is even more plausible to assume the currents divergenceless, so that again they must be completely described by some time-varying tangle of vortex lines at temperatures above any superfluid transition Tc. The observations thus imply that this system, too, is a vortex fluid.

From the observations in references [1] and [2] we may deduce the properties of this vortex fluid state. First, at least in most of the range of observation it is dissipative; the random motions of the vortices constitute a thermal reservoir into which energy may be dissipated, and the current-current correlations decay with time. But it is *incompressible* in the sense that inserting an extra quantum of vorticity costs an energy which is divergent in the distance between such extra vortices. Standard theories of vortex-mediated phase transitions such as Kosterlitz-Thouless[6] in 2D or G A Williams [7] in 3D discuss only the question of adding or removing vortices in opposite sign pairs (ref 6) or vortex loops (ref 7) but we here discuss the addition of net vorticity, and the experiments tell us, rather unequivocally,
 that the response of a vortex liquid to this is anomalous.

Let us make some remarks about the experiments on helium. Several groups have been successful in reproducing results like Chan's using an annular cell, and have found similar ω-dependence of the moment of inertia, in the range where one may estimate that the annular thickness may contain one or a few vortices. Reppy has shown that the amount of NLRS is very dependent on crystal perfection, and increases under conditions where one knows that there are many defects; but Chan's study of very carefully grown samples has been unable to eliminate the effect entirely, and his and Kojima's observations[8] have even hinted that there may be a true phase transition at low enough temperature. But whether there is a true transition in pure He or not, the first message of this paper is that the experimental situation in solid He has been mischaracterized, and that most of the measurements are not consistent with the observation of supersolidity but are with the idea that solid helium contains an incompressible vortex liquid which may be above a supersolidity transition.

We will now try to make the existence of this state theoretically plausible. To do so I will revert to the 2D model of ref [6], although I believe that the results generalize simply to 3D. The current in a 2D system of vortices is simply the sum of those due to the individual vortex points: (We scale $\rho_s$ to 1 for convenience.)

$$J_i = \nabla \Phi_i = q_i \hat{\theta}_i / |r - r_i|$$
$$q_i = \pm 1 \qquad [2]$$

There must be a lower cutoff a around the vortex points if only because the velocity can't be infinite; this will be implicit in all further work. The energy is then the integral of the square of the sum of all the contributions [2]:

$$U = \frac{1}{2} \int d^2 r (\sum_i J_i)^2 \qquad [3]$$

The integration in [3] may be carried out and the result is, introducing an upper cutoff radius R for the sample as a whole which is more or less identical for all vortices:

$$U/2\pi = \sum_i q_i^2 \ln(R/a) + \sum_{i \neq j} q_i q_j \ln R/r_{ij}$$

$$= (\sum_i q_i)^2 \ln R/a - \sum_{i \neq j} q_i q_j \ln r_{ij}/a \qquad [4]$$

We note from the first line of [4] that each individual vortex has a self-energy which diverges logarithmically as $2\pi \ln(R/a)$; but that if the system of vortices is neutral with $\Sigma_i q_i = 0$, the dependence on sample size cancels against terms from the sum of all the other vortices and the standard Kosterlitz-Thouless interaction energy results, with no dependence on the upper cutoff radius:

$$U_{K-T} = -2\pi \sum_{i \neq j} q_i q_j \ln(r_i - r_j)/a \quad + \sum_j E_c \quad [5]$$

The core energy $E_c$ for the local energy cost of a point zero of the boson field could formally be subsumed into a.

If, however, there is a mismatch in the + and - vortex numbers, there remains a divergent term proportional to the logarithm of the upper cutoff radius. This term is formally proportional to the square of the mismatch but there are also long-range uncompensated terms in the interaction [5], and taking these into account it turns out that effectively we must add in a large self-energy $2\pi \ln(R_c/a)$ for each unpaired vortex, if their distribution is reasonably uniform. $R_c$ is approximately the distance between unpaired "field" vortices. **This term has been omitted in all previous treatments of the "normal" bose fluid, as well as in discussions of the superconducting vortex fluid.**[9]

A mismatch in vortex numbers means that the sample has net vorticity, i e is rotating as a whole (or, in the superconducting case, that it is experiencing an external B-field). As has been understood since the '50's[10], the minimum energy configuration will be a uniform array of vortices, which is the closest mimic of rigid rotation. At length scales greater than the distance between unmatched vortices $R_c$ (this will be the magnetic length, $l_B$, in the superconducting case) the physics is macroscopic and classical, and the quantization of vorticity is irrelevant. In this regime the divergent self-energy for r>the lattice constant of the array may be cancelled against whatever source of energy is causing the rotation or against the source energy of the B-field. But there still remains the energy caused by quantization of the vorticity, leading to a nonuniform local velocity. This energy is (if the density of extra vortices is $n_V$) proportional to

$$n_V \ln(R_c^2/a^2) = n_V \ln(1/n_V a^2) \quad [6]$$

$n_V$ is constrained by the need for canceling the divergent terms to be proportional to B for the superconductor and to $\omega$ for the superfluid, as explained above.*

The crucial point which makes the vortex liquid incompressible is that the energy [6] *is not screened out by the thermally excited pairs above Tc.* This is counterintuitive relative to one's experience with the apparently similar system of electrically charged particles; but it is true. This may be understood by simply examining the energy expression [4]. This consists of the sum of all interaction energies, each term proportional to $\ln(r_{ij}/a)$; and the self-energy term, which is independent of all $r_{ij}$'s and depends on lnR. Adding an unmatching vortex cannot change the interaction sum by this much: at a distance, the extra vortex encounters a neutral gas of vortices, and close by, it may have attracted a screening cloud—consisting of a single quantum of vorticity-- at a radius of order $1/(n_{pairs})^{1/2}$, but that doesn't give a large term in the energy. When a quantum of vorticity is added, its circulating current is uncorrelated with the others' at distances less than $R_c$. We may think of this temperature region as being dominated by entropy, leaving the vortices quite uncorrelated but uniformly distributed.

---

* Some have found this discussion obscure. I have tried to make it parallel between the bose superfluid and the charged pair fluid, where one can remove the divergence in R with a vector potential A. In both cases, however, the limiting scale at large distances is set by the average spacing of the extra "field" vortices. Perhaps it is easiest for the reader to mentally subdivide the system into cells, each containing one "field" vortex, for which the argument is obvious.

Another way to understand this rather counterintuitive point is to realise that the change in the interaction sum [5] will not depend on whether the added vortex is the first or the second of a ± pair, the screening dynamics will be basically local.  Since [5] is the only part of the energy depending on relative vortex positions, that leaves the term [6] unchanged.

In 3 dimensions, the computation is more complex but if we are above the x-y model Tc, entropy dominates and the extra vortex line is not effectively screened.

On the other hand,  [4] is the energy, not the free energy.  Below Tc, it is controlling and gives us, for instance, the Abrikosov theory of the vortex lattice in superconductors. Below Tc the thermally excited vortices are bound in pairs and partially screen the interactions.  In the Kosterlitz-Thouless theory, Tc occurs where the extra logarithmic energy of free vortices is compensated in the free energy by T times the logarithmic entropy which one gains by allowing the vortex to be anywhere in the sample.  Above Tc, pairs of vortices proliferate in such a way that their number is given by the activation expression which results from equating the logarithmic terns in the energy and entropy of a nearby pair:

$$n_{pr} = (1/a^2)\exp[-E_c/(T-T_c)] \qquad [7]$$

Interactions reduce the numbers of vortices but do not cause strong correlations among them.

The large extra entropy of the unmatched vortices does not cause them to proliferate because they are indistinguishable from the positive members of pairs whose number obeys [7]. Actually, the free energy of order nln(n) cancels exactly. This cancellation does not occur to higher order in $n_V$ leading to a free energy term of form $(n_V)^2 \ln(1/n_V)$, which gives the logarithmically divergent response function.

The Nernst effect gives us a uniquely direct way of measuring specifically the energy carried by vortices. This is because of the reciprocity between the Nernst effect—the voltage response to a heat current—and the Ettingshausen effect, the heat current response to a given voltage. An E field implies that the net vorticity moves at the velocity

$$v = cE/B \qquad [8]$$

The heat transported is just the actual energy [6] of the vortices, so the Ettingshausen coefficient $\alpha_{xy}$ is proportional to $B \ln B$. The actual shape of $\alpha_{xy}$ vs B is complicated by the strong dependence of $\rho_s$ on B, and has been discussed at length elsewhere.[11]

An equally challenging experiment in the superconducting case is the direct measurement of the vortex energy via the magnetization. The energy must come from the interaction of the current with the field, which is J·A/2 or, equivalently, M·B/2, so the diamagnetic moment is a direct measurement of the energy due to the added vortices. When, as is often the case, the two measurements yield nearly identical results related by the factor 2/T, that is strong evidence for quantized vortices.

Both the Nernst effect and the magnetization can have other causes. There are several ways in which a Nernst effect can result from particle currents, although except in special circumstances these effects are small and linear. It is also clear that magnetization can result from spin susceptibility or Van Vleck paramagnetism, which are completely independent effects. Thus identifying the vortex term by using the two measurements together may often allow a fairly unequivocal diagnosis of vortex motion.

Most treatments of the Bose liquid above Tc such as ref 9 have restricted themselves to the critical range near the λ point or K-T transition. But as we see, the anomalous response is not a critical phenomenon but an intrinsic property of the vortex liquid phase, and should persist as long as there is a finite core energy for vortices. In Ong's Nernst effect fluid there seems to be quite a range above the critical region which is characterized by a correlation time for vortex flow of around h/kT, which then sets the density of vortices via v=h/m($\nabla\phi$). The vortex liquid phase seems to die away slowly and continuously in solid He, according to Kubota (Private communication.)

Acknowledgements: extensive discussions with D A Huse, N P Ong, Yayu Wang, W F Brinkman, V Oganesyan, Sri Ragu, M Chan, J. Reppy, H Kojima, and M Kubota.